\documentstyle[12pt]{article}
\def\singlespace {\smallskipamount=3.75pt plus1pt minus1pt
                  \medskipamount=7.5pt plus2pt minus2pt
                  \bigskipamount=15pt plus4pt minus4pt
                  \normalbaselineskip=15pt plus0pt minus0pt
                  \normallineskip=1pt
                  \normallineskiplimit=0pt
                  \jot=3.75pt
                  {\def\smallskip {\vskip\smallskipamount}}
                  {\def\medskip   {\vskip\medskipamount}}
                  {\def\bigskip   {\vskip\bigskipamount}}
                  {\setbox\strutbox=\hbox{\vrule
                    height10.5pt depth4.5pt width 0pt}}
                  \parskip 7.5pt
                  \normalbaselines}
\def\middlespace {\smallskipamount=5.625pt plus1.5pt minus1.5pt
                  \medskipamount=11.25pt plus3pt minus3pt
                  \bigskipamount=22.5pt plus6pt minus6pt
                  \normalbaselineskip=22.5pt plus0pt minus0pt
                  \normallineskip=1pt
                  \normallineskiplimit=0pt
                  \jot=5.625pt
                  {\def\smallskip {\vskip\smallskipamount}}
                  {\def\medskip   {\vskip\medskipamount}}
                  {\def\bigskip   {\vskip\bigskipamount}}
                  {\setbox\strutbox=\hbox{\vrule
                    height15.75pt depth6.75pt width 0pt}}
                  \parskip 11.25pt
                  \normalbaselines}
\def\doublespace {\smallskipamount=7.5pt plus2pt minus2pt
                  \medskipamount=15pt plus4pt minus4pt
                  \bigskipamount=30pt plus8pt minus8pt
                  \normalbaselineskip=30pt plus0pt minus0pt
                  \normallineskip=2pt
                  \normallineskiplimit=0pt
                  \jot=7.5pt
                  {\def\smallskip {\vskip\smallskipamount}}
                  {\def\medskip   {\vskip\medskipamount}}
                  {\def\bigskip   {\vskip\bigskipamount}}
                  {\setbox\strutbox=\hbox{\vrule
                    height21.0pt depth9.0pt width 0pt}}
                  \parskip 15.0pt
                  \normalbaselines}
\begin{document}

\newcount\sectionnumber
\sectionnumber=0

\title{Exact Relations for Heavy-Light Quark Systems}

\author{
Patrick J. O'Donnell \\
Physics Department,\\
University of Toronto,\\
Toronto, Ontario M5S 1A7, Canada.\\
\\
and \\
\\
Humphrey K.K. Tung\\
Institute of Physics,\\
Academia Sinica,\\
Taipei, Taiwan 11529, R.O.C.
}

\date{UTPT-93-16}

\maketitle

\begin{abstract}
\middlespace

We derive general relations among hadronic form factors involving one
heavy meson $b\overline{q}$ and another, not necessarily heavy, meson
$Q\overline{q}$.
The relations are valid to all orders of mass corrections of $m_Q$ .

\end{abstract}

\newpage
\middlespace

The inclusive rare decay $B\rightarrow X_{s}\gamma$ is now well
understood in the context of the standard model \cite{stan} and the
new experimental upper bound \cite{CLEO1} of $5.4\times 10^{-4}$ for the
branching ratio is already playing an important role \cite{Hew} in
constraining the parameters of various models other than the standard model.
The first experimental observation of the exclusive decay
$B\rightarrow K^{\ast}\gamma$ has been reported from the
CLEO collaboration \cite{CLEO2} which gives a branching ratio for
this mode of $(4.5 \pm 1.5 \pm 0.9)\times 10^{-5}$.
It is this exclusive rare decay
$B\rightarrow K^{\ast}\gamma$, however, which
is the least well known theoretically due to the large
recoil momentum of the $K^{\ast}$ meson \cite{OT}.
We have recently shown \cite{OT1} that heavy-quark symmetry
together with $SU(3)$ flavor symmetry could relate the rare decay
$B\rightarrow K^{\ast}\gamma$ to a measurement of
the semileptonic decay $B\rightarrow\rho e\bar{\nu}$. This work made
use of the heavy-quark limit for the $B$ meson and the weak binding
limit for the $K^{\ast}$ meson. However, we also showed that the results
were not an artifact of a particular quark model by demonstrating that the
agreement of the form factor relations also held for the
BSW model \cite{BSW}.

In analysing the relations among the form factors, we noted that there were
a number of relations that only depend on the heavy-quark limit for the $b$
quark and not on the weak binding limit. It is these relations
which we describe here.
Although the interest is in the $K^{\ast}$ meson, we shall describe the
decays from
a $B(b\bar{q})$
or $B^{\ast}$ to an arbitrary vector meson $V(Q\overline q)$, so that
our results are also applicable to the charm system.

First, we define the hadronic form factors of interest as
in Ref. \cite{FN}:-

\vbox{
\begin{eqnarray}
\langle V(v^{\prime},\epsilon)|\bar{Q}\gamma_{\mu}b|B(v) \rangle
&=& \sqrt{m_{B}m_{V}} \, h_{V}
i\varepsilon_{\mu\nu\lambda\sigma}\epsilon^{\ast\nu}
v^{\prime\lambda}v^{\sigma}\,\, ,\\
\langle V(v^{\prime},\epsilon)|\bar{Q}\gamma_{\mu}\gamma_{5}b|B(v) \rangle
&=& \sqrt{m_{B}m_{V}} \left[ \,
h_{A_{1}}(1+w)\epsilon^{\ast}_{\mu}
- h_{A_{2}}(\epsilon^{\ast}\cdot v)v_{\mu}
- h_{A_{3}}(\epsilon^{\ast}\cdot v)v^{\prime}_{\mu}\, \right] \,\, ,\\
\langle V(v^{\prime},\epsilon)|\bar{Q}\gamma_{\mu}b|B^{\ast}(v,\zeta)\rangle
&=& \sqrt{m_{B}m_{V}} \left\{ \rule{0cm}{0.2cm} \,
-(\zeta\cdot\epsilon^{\ast}) \left[ \,h_{1}(v+v^{\prime})_{\mu}
+ h_{2}(v-v^{\prime})_{\mu} \, \right] \right. \nonumber\\
&+& \left. h_{3}(\epsilon^{\ast}\cdot v)\zeta_{\mu}
+ h_{4}(\zeta\cdot v^{\prime} )\epsilon^{\ast}_{\mu}
- (\zeta\cdot v^{\prime})(\epsilon^{\ast}\cdot v)
[ \, h_{5}v_{\mu}+h_{6}v^{\prime}_{\mu} \, ] \,
\rule{0cm}{0.2cm} \right\} \, , \\
\langle V(v^{\prime},\epsilon)|\bar{Q}\gamma_{\mu}\gamma_{5}b
|B^{\ast}(v,\zeta)\rangle
&=& \sqrt{m_{B}m_{V}}\, i\varepsilon_{\mu\nu\lambda\sigma} \left\{ \,
\zeta^{\lambda}\epsilon^{\ast\sigma}
\left[ \, h_{7}(v+v^{\prime})^{\nu}
+ h_{8}(v-v^{\prime})^{\nu} \, \right] \right.\nonumber \\
&+& \left.  v^{\prime\lambda}v^{\sigma} \left[
\, h_{9}(\epsilon^{\ast}\cdot v)\zeta^{\nu}
+ h_{10}(\zeta\cdot v^{\prime}) \epsilon^{\ast\nu}\, \right] \,\right\}
\,\,\, .\label{hs}
\end{eqnarray}}
The terms $\zeta$ and $\epsilon^{\ast}$ are the polarization vectors
of $B^{\ast}$ and $V$, respectively. The variables $v$ and
$v^{\prime}$ are the four velocities of $B^{\ast}$ (or $B$) and
$V$, and we define $w=v\cdot v^{\prime}$. We show below that the
form factors for the decays $B\rightarrow V$ and $B^{\ast}\rightarrow V$
can be related using the spin symmetry and static limit of the
heavy $b$ quark.

In the heavy $b$ limit, the spin of the $b$ quark is
decoupled from all other light fields in the $B$ meson \cite{IW1}.
We can therefore construct the spin operator $S^{Z}_{b}$ for the $b$
quark such that
\[
S^{Z}_{b}|B(b\bar{q})\rangle
= \frac{1}{2}|B^{\ast}_{l}(b\bar{q})\rangle\,\,\, ,
\,\,\, S^{Z}_{b}|B^{\ast}_{l}(b\bar{q})\rangle
= \frac{1}{2}|B(b\bar{q})\rangle\,\,\, ,
\]
where $B^{\ast}_{l}$ stands for a longitudinal vector $B^{\ast}$ meson.
In $|B\rangle$ and $|B^{\ast}_{l}\rangle$,
the spatial momentum of the $b$ quark is defined in
the $z$-direction for the $b$ spinor to be an eigenstate
of $S^{Z}_{b}$. Using the relation
$\langle V|\,\bar{Q}\Gamma b\,|B\rangle =
-2\langle V|\,[S^{Z}_{b},\bar{Q}\Gamma b]\,|B^{\ast}_{l}\rangle $,
for $\Gamma$ any product of $\gamma$ matrices,
we have the following identities between the
$B\rightarrow V$ and $B^{\ast}_{l}\rightarrow V$ matrix elements:
\begin{eqnarray}
\langle V|A_{0}|B\rangle &=& -\langle V|V_{3}|B^{\ast}_{l}
\rangle\,\,\, ,\label{v1}\\
\langle V|A_{3}|B\rangle &=& -\langle V|V_{0}|B^{\ast}_{l}
\rangle\,\,\, ,\label{v2}\\
\langle V|V_{\pm}|B\rangle &=& \mp\langle V|V_{\pm}|B^{\ast}_{l}
\rangle\,\,\, ,\label{v3}\\
\langle V|V_{0}|B\rangle &=& -\langle V|A_{3}|B^{\ast}_{l}
\rangle\,\,\, ,\label{a1}\\
\langle V|V_{3}|B\rangle &=& -\langle V|A_{0}|B^{\ast}_{l}
\rangle\,\,\, ,\label{a2}\\
\langle V|A_{\pm}|B\rangle &=& \mp\langle V|A_{\pm}|B^{\ast}_{l}
\rangle\,\,\, ,\label{a3}
\end{eqnarray}
where $V_{\mu}=\bar{Q}\gamma_{\mu}b$ and
$A_{\mu}=\bar{Q}\gamma_{\mu}\gamma_{5}b$.

Using the matrix identities in Eqs. (\ref{v1}-\ref{a3}), we can relate
the form factors of $B\rightarrow V$ to those of
$B^{\ast}\rightarrow V$. Since the spatial
momentum of the $b$ quark is defined in the $z$-direction, we should
work in the $B$ rest frame and choose the longitudinal polarization
vector for $B^{\ast}_{l}$ to be $\zeta^{\mu}_{l}=(0;0,0,1)$.
The matrix identities are evaluated for both transverse and
longitudinal polarizations of the vector meson $V(Q\overline q)$.
This gives the following relations among the form factors:-
\begin{eqnarray}
h_{4} &=& h_{1}-h_{2}\,\,\, ,\nonumber\\
h_{5} &=& h_{9}\,\,\, ,\nonumber\\
h_{6} &=& 0\,\,\, ,\nonumber\\
h_{7} &=& h_{1}\,\,\, ,\nonumber\\
h_{8} &=& h_{2}\,\,\, ,\nonumber\\
h_{10} &=& 0 \,\,\, ,\nonumber\\
h_{V} &=& h_{1}-h_{2}\,\,\, ,\nonumber\\
h_{A_{1}} &=& (h_{1}-h_{2})+\frac{2h_{2}}{(1+w)}\,\,\, ,\nonumber\\
h_{A_{2}} &=& (h_{1}+h_{2}-h_{3})+wh_{9}\,\,\, ,\nonumber\\
h_{A_{3}} &=& (h_{1}-h_{2})-h_{9}\,\,\, .
\label{rh}
\end{eqnarray}
In the recent literature \cite{FN,eff} these relations are obtained using an
effective Lagrangian approach. While this eases the burden in calculating
the symmetry limit when both mesons contain heavy constituent quarks
(and allows for a systematic inclusion of mass corrections \cite{FN}),
we shall see that Eqs. (\ref{v1} - \ref{a3}) are simpler to use when
the symmetry may be badly broken, or even is not applicable.

In the decay $B\rightarrow K^{\ast}\gamma$ in particular, we consider
also the hadronic matrix element for the current
$\bar{Q}i\sigma_{\mu\nu}q^{\nu}b_{R}$, where $q=p_{B}-k$ is the
momentum of the outgoing photon. The covariant expansion of the
matrix element is given by
\begin{eqnarray}
\langle V(k,\epsilon)|\bar{Q}i\sigma_{\mu\nu}q^{\nu}b_{R}|B(p_{B})\rangle
&=& f_{1}(q^{2})i\varepsilon_{\mu\nu\lambda\sigma}
\epsilon^{\ast\nu}p^{\lambda}_{B} k^{\sigma}\nonumber\\
&& + \left[(m^{2}_{B}-m^{2}_{V})\epsilon^{\ast}_{\mu}-
(\epsilon^{\ast}\cdot q)(p_{B}+k)_{\mu}\right]f_{2}(q^{2})\nonumber\\
&& + (\epsilon^{\ast}\cdot q)\left[(p_{B}-k)_{\mu}-\frac{q^{2}}{(m^{2}_{B}
-m^{2}_{V})}(p_{B}+k)_{\mu}\right]f_{3}(q^{2})\, .
\end{eqnarray}
We can relate the form factors $f_{1,2,3}$ to $h$s defined in Eq. (\ref{hs})
using the static limit of the $b$ quark. In the $B$ rest frame,
the static $b$-quark spinor satisfies the equation of motion $\gamma_{0}b=b$.
We then have the relations between the $\gamma_{\mu}$
and $\sigma_{\mu\nu}$ matrix elements \cite{IW2}:-
\begin{eqnarray}
\langle V | \bar{Q}\gamma_{i}b|B\rangle &=&
\langle V | \bar{Q}i\sigma_{0i}b|B\rangle\,\,\, ,
\\
\langle V | \bar{Q}\gamma_{i}\gamma_{5}b|B\rangle
&=& - \langle V | \bar{Q}i\sigma_{0i}\gamma_{5}b|B
\rangle\,\,\, .
\end{eqnarray}
This gives the form-factor relations
\begin{eqnarray}
h_{f_{1}} &=&(m_{B}+m_{V})(h_{1}-h_{2})+2m_{V}h_{2}\,\,\, ,\nonumber\\
h_{f_{2}} &=& \frac{m_{B}m_{V}}{(m_{B}+m_{V})}(1+w)(h_{1}-h_{2})
+ \frac{(m_{B}-wm_{V})}{(m^{2}_{B}-m^{2}_{V})}
2m_{B}m_{V}h_{2}\,\,\, ,\nonumber\\
h_{f_{3}} &=& \frac{1}{2}(m_{B}-m_{V})(h_{1}-h_{2})
-m_{V}h_{2} -\frac{(m^{2}_{B}-m^{2}_{V})}{2m_{B}}h_{9}\,\,\, ,
\label{rf}
\end{eqnarray}
where $h_{f_{1}}=\sqrt{4m_{B}m_{V}}f_{1}$,
$h_{f_{2}}=\sqrt{4m_{B}m_{V}}f_{2}$, and
$h_{f_{3}}=\sqrt{4m_{B}m_{V}}f_{3}$.
Thus, using only the spin symmetry and static limit of the heavy
$b$ quark, we can express the $B\rightarrow V$ and
$B^{\ast}\rightarrow V$ hadronic form factors in terms of
four independent form factor combinations
$h_{1}-h_{2}$, $h_{2}$, $h_{1}+h_{2}-h_{3}$, and $h_{9}$
as shown in Eqs. (\ref{rh}) and (\ref{rf}).
Note that we have made no assumptions about the mass of the quark $Q$
at any point in the above discussion.
The form factor relations are therefore valid for {\em heavy} or
{\em light} $Q$.

In terms of the factors $\epsilon$, $\bar{\epsilon}$,
and $\rho$ defined as
\begin{eqnarray}
- \frac{(h_{1}+h_{2}-h_{3})}{(h_{1}-h_{2})}
&\equiv & (1+w) \bar{\varepsilon} \,\,\, ,\nonumber\\
- \frac{2h_{2}}{(h_{1}-h_{2})} &\equiv & (1+w)\varepsilon \,\,\, ,
\nonumber\\
- \frac{h_{9}}{(h_{1}-h_{2})} &\equiv & \rho \,\,\, ,\nonumber
\end{eqnarray}
we can rewrite the form factor relations in Eqs. (\ref{rh}) and
(\ref{rf}) in terms of $h_{V}=h_{1}-h_{2}$ and the three newly
defined factors as,
\begin{eqnarray}
h_{3} &=& h_{V} [ 1 + (1+w)(\bar{\varepsilon}-\varepsilon ) ]
\,\,\, ,\nonumber\\
h_{4} &=& h_{V}\,\,\, ,\nonumber\\
h_{9} &=& h_{5} = -\rho h_{V}\,\,\, ,\nonumber\\
h_{6} &=& 0 \,\,\, ,\nonumber\\
h_{1} &=& h_{7} = h_{V}\left( 1- \frac{1+w}{2}\varepsilon \right) \,\,\, ,
\nonumber\\
h_{2} &=& h_{8} =  - \frac{1+w}{2}\varepsilon h_{V}\,\,\, ,\nonumber\\
h_{10} &=& 0 \,\,\, ,\nonumber\\
h_{A_{1}} &=& h_{V}( 1-\varepsilon )\,\,\, ,\nonumber\\
h_{A_{2}} &=& - h_{V} \left[ \, (1+w)\bar{\varepsilon} + w\rho \, \right]
\,\,\, ,\nonumber\\
h_{A_{3}} &=& h_{V} ( 1+\rho )\,\,\, ,\nonumber\\
h_{f_{1}} &=& h_{V} \left[ \, (m_{B}+m_{V}) - m_{V}(1+w)\varepsilon
\, \right] \,\,\, ,\nonumber\\
h_{f_{2}} &=& h_{V}\, m_{B}m_{V} (1+w) \left[ \, \frac{1}{(m_{B}+m_{V})}
- \frac{ (m_{B}-wm_{V}) }{(m^{2}_{B}-m^{2}_{V})}\varepsilon
\, \right] \,\,\, ,\nonumber\\
h_{f_{3}} &=& h_{V} \left[ \, \frac{1}{2}(m_{B}-m_{V})
+ \frac{1}{2}m_{V}(1+w)\varepsilon +
\frac{(m^{2}_{B}-m^{2}_{V})}{2m_{B}}\rho \,\right] \,\,\, .
\label{FNEQ2}
\end{eqnarray}
In the heavy $Q$ limit \cite{FN}, we have $h_{1}=h_{3}$ and
$h_{2}=h_{9}=0$; thus, the factors $\varepsilon$, $\bar{\varepsilon}$,
and $\rho$ all vanish in this limit. For finite quark mass $m_{Q}$,
however, they should represent the full $1/m_{Q}$ corrections to
the heavy-quark symmetry relations.
By simple inspection we see in particular that the symmetry relations
in Eq. (\ref{FNEQ2}) are consistent
with the order $1/m^{2}_Q$ result in Ref. \cite{FN}.
For the form factors $h_{A_1}$, $h_{1}$ and
$h_{7}$ we can see that at the point where Luke's \cite{Luke} theorem
would set in, {\it i.e.} $\omega=1$ and for heavy $m_b$ and $m_Q$, the
renormalization is the same to all orders in the mass of $Q$.
(By inspection, the same results holds true for all orders in
the $b$ quark for large $Q$ but the mixed expansion of the masses is
different).

We can estimate the size of the correction factors
$\epsilon$, $\bar{\epsilon}$, and $\rho$ using the nonrelativistic
quark model \cite{ISGW}. In the quark model, we have in
the $B$ rest frame
\begin{equation}
- \rho \approx \bar{\epsilon} = \epsilon =
1 - (E_{V}-m_{V})\frac{H_{1}}{H_{2}} \,\,\, , \label{QM}
\end{equation}
where $E_{V}=(m^{2}_{B}+m^{2}_{V}-q^{2})/(2m_{B})$ is the energy
of $V$. The terms $H_{1}$ and $H_{2}$ are overlapping
integrals of the momentum wave functions in the quark model,
they are given by
Eq. (22) of our recent paper \cite{OT1}.
In Ref. \cite{OT1}, we have shown that in the weak binding
limit of $V$ the factor
$1-(E_{V}-m_{V})H_{1}/H_{2}$ in Eq. (\ref{QM}) is much smaller than 1
throughout the whole kinematic region.
In an numerical calculation using the ISGW
parameterization \cite{ISGW} of the momentum wave function,
we show explicitly that
$1-(E_{V}-m_{V})H_{1}/H_{2} < 0.05$ for $B\rightarrow K^{\ast}$
and $ 0.11 $ for $B\rightarrow\rho$ throughout the full
kinematic range. Accordingly, the factors $\epsilon$,
$\bar{\epsilon}$, and $\rho$ can be treated as small corrections
to the heavy-quark symmetry relations even for light $Q$.

It is not surprising that the above result
coincides with the one that emerges in the
heavy-quark limit, since the spin symmetry for $Q$ should
approximately hold in the quark model.
While the validity of estimating the
size of correction factors using the nonrelativistic quark model may
be questioned,
we will show that the same result can be obtained using
a relativistic quark model.
In the BSW model \cite{BSW}, we have at the maximum recoil of $V$
($q^{2}=0$ or $v\cdot v^{\prime}=(m^{2}_{B}+m^{2}_{V})/(2m_{B}m_{V})$)
the expressions for $\varepsilon$, $\bar{\varepsilon}$,
and $\rho$ given by
\begin{eqnarray}
\varepsilon (0) &=& 1 - \left(\frac{ m_{B}-m_{V}}{m_{B}+m_{V}} \right)
\left( \frac{m_{b}+m_{Q}}{m_{b}-m_{Q}} \right)\,\,\, ,\\
\bar{\varepsilon} (0) &=& \frac{1}{(m_{b}-m_{Q})}
\frac{4m^{2}_{B}m_{V}}{(m_{B}+m_{V})^{2}}
\left[ \, \frac{g_{1}}{g_{2}}
- \frac{1}{2} \left( \frac{m_{b}}{m_{B}} + \frac{m_{Q}}{m_{V}} \right)
\left( 1 + \frac{m_{V}}{m_{B}} \right) \,\right] \,\,\, ,\\
\rho (0) &=& \frac{1}{(m_{b}-m_{Q})}
\frac{-2m_{B}m_{V}}{(m_{B}-m_{V})}
\left[ \, \frac{m_{b}}{m_{B}} - \frac{m_{Q}}{m_{V}}
\left( \frac{m_{B}-m_{V}}{m_{B}+m_{V}} \right) \,\right]\,\,\, .
\end{eqnarray}
The terms $g_{1}$ and $g_{2}$ are overlap integrals given by
\begin{eqnarray}
g_{1} &=& \int d{\bf p}_{T} \int^{1}_{0} dx
\phi^{\ast}_{V}({\bf p}_{T},x)\phi_{B}({\bf p}_{T},x)\,\,\, ,\\
g_{2} &=& \int d{\bf p}_{T} \int^{1}_{0} \frac{dx}{x}
\phi^{\ast}_{V}({\bf p}_{T},x)\phi_{B}({\bf p}_{T},x)\,\,\, .
\end{eqnarray}
In the BSW model, the orbital wave function $\phi_{B}$ and
$\phi_{V}$ are solutions to a relativistic scalar harmonic
oscillator potential. Notice that only
$\bar{\varepsilon}(0)$ depends on the overlapping effects
in the BSW model, and $\bar{\varepsilon}$ enters only into
the expressions for $h_{A_{2}}$ and $h_{3}$ in the symmetry relations
of Eq. (\ref{FNEQ2}). All other relations in (\ref{FNEQ2})
are overlap independent. It can be shown that the ratio
$g_{1}/g_{2}$ is very stable with respect to the parameter changes
in the orbital wave functions in the BSW model.
Numerically, for the decay $B\rightarrow K^{\ast}$,
we have the values of the correction factors given by
$\bar{\varepsilon }^{B\rightarrow K^{\ast}}(0)= -0.023 $,
$\varepsilon^{B\rightarrow K^{\ast}}(0)= 0.11 $, and
$\rho^{B\rightarrow K^{\ast}}(0)=-0.25$.
For the decay $B\rightarrow \rho$, we have
$\bar{\varepsilon}^{B\rightarrow \rho}(0)= 0.042 $,
$\varepsilon^{B\rightarrow \rho}(0)= 0.15 $, and
$\rho^{B\rightarrow \rho}(0)=-0.24$.

In a recent paper \cite{OT1}, we have discussed a method of
relating the
decay $B\rightarrow K^{\ast}\gamma$ to the semileptonic decay
$B\rightarrow\rho e\bar{\nu}$ using $SU(3)$ \cite{BD}
symmetry and the symmetry relations in Eq. (\ref{FNEQ2}).
The ratio of the decay $B\rightarrow K^{\ast}\gamma$ to the
semileptonic decay
$B\rightarrow\rho e\bar{\nu}$ is shown to be proportional to
the factor ${\cal I}$ which is equal to 1 in the heavy-quark limit.
For finite quark mass $m_{Q}$, the factor ${\cal I}$ takes
into account the corrections to
the symmetry relations coming from $\epsilon$, $\bar{\epsilon}$,
and $\rho$. In Ref. \cite{OT1}, however, the hadronic matrix element
$\langle V|\bar{Q}\gamma_{\mu}\gamma_{5}b|B^{\ast}\rangle $
does not include the $h_{9}$ and $h_{10}$ terms as in Eq. (\ref{hs}).
Including the form factors $h_{9}$ and $h_{10}$,
we have the correct expression for ${\cal I}$ given by
\begin{equation}
{\cal I} = \frac{ 1 - \frac{(m_{B}+m_{K^{\ast}})}{2m_{B}}
\varepsilon^{B\rightarrow K^{\ast}}(0) }
{ 1
+ \frac{(m_{B}-m_{\rho})(m_{B}+m_{\rho})^{2}}{4m^{2}_{B}m_{\rho}}
\bar{\varepsilon}^{B\rightarrow\rho}(0)
- \frac{(m_{B}+m_{\rho})}{2m_{\rho}}
\varepsilon^{B\rightarrow\rho}(0)
- \frac{(m_{B}-m_{\rho})^{2}(m_{B}+m_{\rho})}{4m^{2}_{B}m_{\rho}}
\rho^{B\rightarrow\rho}(0) } \,\,\, .
\end{equation}
and the value of ${\cal I}=1.12$ which remains close to 1.

\vspace{.3in} \centerline{ {\bf Acknowledgment}}

This work was supported by the Natural Sciences and Engineering
Council of Canada and by the National Sciences
Council of the Republic of China.

\newpage
\middlespace

\end{document}